\documentclass[sigconf]{acmart}
\usepackage{xcolor}
\usepackage[linesnumbered, ruled, vlined]{algorithm2e}

\SetCommentSty{mycommentstyle}

\usepackage{tikz}
\usepackage[linesnumbered, ruled, vlined]{algorithm2e}
\usepackage{amsmath}
\usepackage{tabularx}
\usepackage{booktabs}
\usepackage{graphicx}
\usepackage{subcaption}
\usepackage{multirow}
\usepackage{enumitem}
\usepackage{float} 

\AtBeginDocument{%
  }

\setcopyright{acmlicensed}
\copyrightyear{2025}
\acmYear{2025}
\acmDOI{XXXXXXX.XXXXXXX}
\acmConference[RecSys '25]{The 19th ACM Conference on Recommender Systems}{September 22--26, 2025}{Prague, Czech Republic}
\acmISBN{978-1-4503-XXXX-X/2025/09}

\begin{document}

\title{Privacy Risks of LLM-Empowered Recommender Systems: \\
An Inversion Attack Perspective}

\author{Yubo Wang}
\authornote{Both authors contributed equally to this research.}
\email{ywan0813@student.monash.edu}
\orcid{0009-0006-6088-9907}
\affiliation{%
  \institution{Monash University}
  \city{Melbourne}
  \state{Victoria}
  \country{Australia}
}

\author{Min Tang}
\authornotemark[1]
\email{min.tang@monash.edu}
\orcid{0009-0000-1088-8523}
\affiliation{%
  \institution{Monash University}
  \city{Melbourne}
  \state{Victoria}
  \country{Australia}
}

\author{Nuo Shen}
\email{nshe0037@student.monash.edu}
\orcid{0009-0001-8363-0125}
\affiliation{%
  \institution{Monash University}
  \city{Melbourne}
  \state{Victoria}
  \country{Australia}
}
 
\author{Shujie Cui}
\authornote{Corresponding author}
\email{Shujie.Cui@monash.edu}
\orcid{0000-0001-8124-6800}
\affiliation{%
  \institution{Monash University}
  \city{Melbourne}
  \state{Victoria}
  \country{Australia}
}

\author{Weiqing Wang}
\email{Teresa.Wang@monash.edu}
\orcid{0000-0002-9578-819X}
\affiliation{%
  \institution{Monash University}
  \city{Melbourne}
  \state{Victoria}
  \country{Australia}
}

\renewcommand{\shortauthors}{Wang et al.}

\begin{abstract}
The large language model (LLM) powered recommendation paradigm has been proposed to address the limitations of traditional recommender systems (RecSys), which often struggle to handle cold-start users or items with new IDs. 
Despite its effectiveness, this study uncovers that LLM-empowered RecSys are vulnerable to reconstruction attacks that can expose both system and user privacy. To thoroughly examine this threat, we present the first systematic study on inversion attacks targeting LLM-empowered RecSys, wherein adversaries attempt to reconstruct original prompts that contain personal preferences, interaction histories, and demographic attributes by exploiting the output logits of recommendation models. 
We reproduce the \texttt{vec2text} framework and optimize it using our proposed method - Similarity-Guided Refinement, enabling more accurate reconstruction of textual prompts from model-generated logits.
Extensive experiments across two domains (movies and books) and two representative LLM-based recommendation models demonstrate that our method achieves high-fidelity reconstructions.  Specifically, we can recover nearly 65\% of the user-interacted items and correctly infer age and gender in 87\% of the cases. The experiments also reveal that privacy leakage is largely insensitive to the victim model’s performance but highly dependent on domain consistency and prompt complexity. These findings expose critical and privacy vulnerabilities in LLM-empowered RecSys.
The code for reproduction is provided below: \url{https://github.com/xuemingxxx/Attack_RecSys/}
\end{abstract}

\ccsdesc[500]{Information systems~Recommender systems}
\ccsdesc[300]{Computing methodologies~Machine learning}

\keywords{Recommender Systems, Large Language Models, Privacy Risks, Model Inversion Attack}

\maketitle

\section{INTRODUCTION}

\begin{figure}[htbp]
    \centering
    \includegraphics[width=0.46\textwidth]{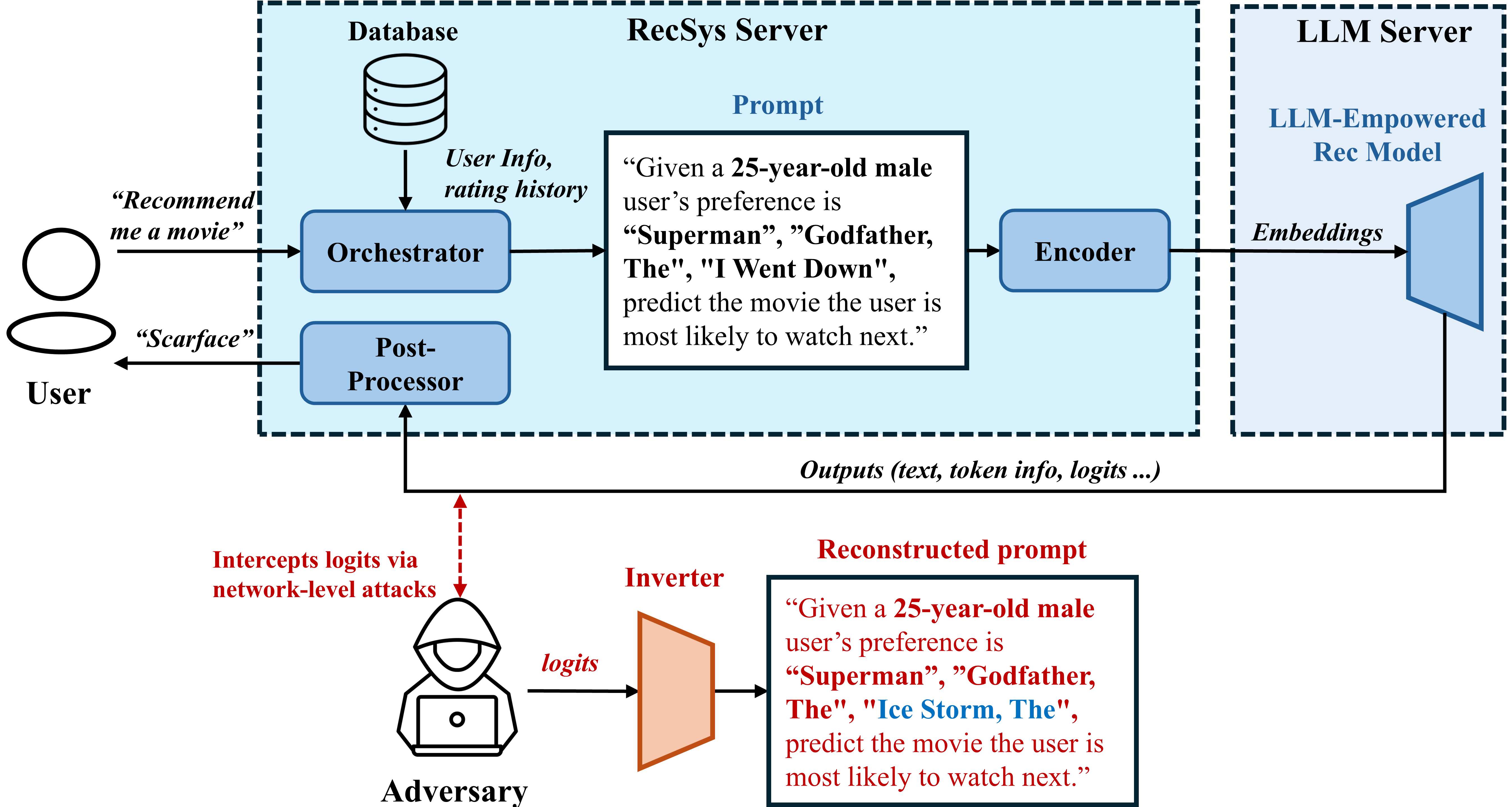} 
    \caption{Inversion Attack against LLM-empowered RecSys: A user interacts with the system via natural language queries, which are internally formatted into a prompt and processed by an LLM-empowered recommendation model. The adversary intercepts the logits returned to the recommender system for post-processing and feeds them into an inverter to reconstruct the original prompt, thereby exposing sensitive user attributes and preferences.
    }
    \label{fig:intro}
\end{figure}

Recommender systems (RecSys) have become essential to modern online platforms, helping guide users to products, content, or services that align with their individual preferences \cite{goyani2020review,ekstrand2011collaborative,liu2025flow}.
Traditionally, RecSys approaches have leaned on collaborative filtering \cite{ekstrand2011collaborative, koren2009matrix, sarwar2001item} or content-based methods \cite{lops2011content,pazzani2007content} to connect users with items they might be interested in. 
More recently, large language models (LLMs) have unlocked a new dimension of personalization for RecSys. 
Rather than depending solely on abstract ID-based data, LLM-powered RecSys reimagines all information through a linguistic framework \cite{bao2023tallrec,geng2022recommendation,zhang2025collm, zhu2024collaborative, wang2024rethinking}. 
This includes system instructions, context, user profiles, and historical interactions, which are combined to prompt the system for more accurate and contextually relevant recommendations.
For example, P5 \cite{geng2022recommendation} leverages LLM’s capabilities to significantly enhance recommendation performance by understanding nuanced user preferences and item descriptions.

The powerful semantic understanding and superior generalization abilities of LLM significantly enhance recommendation performance, particularly when addressing the cold-start problem \cite{zhang2025coldstart, sanner2023large,kim2024large, tang2025sequential}.
However, the integration of LLMs also introduces potential vulnerabilities.
For example, CheatAgent \cite{zhang2024stealthy} can subtly increase an item’s exposure by merely altering its textual content during the testing phase. 
Similarly, Ning et al. \cite{ning2024cheatagent, ning2025exploring} develop an agent that generates adversarial perturbations on input prompts and demonstrate that poisoning merely 1\% of the training data is sufficient to mislead LLM-empowered RecSys into making incorrect decisions. 
While these approaches typically aim to undermine recommendation accuracy or manipulate the promotion of specific items, they overlook the privacy vulnerabilities inherent in LLM-empowered RecSys.

Recent studies \cite{morris2023language, chen2025algen, zhuang2024does, zhang2024extracting} reveal that general LLMs are vulnerable to inversion attacks, where adversaries reconstruct input prompts from output text or next-token probabilities (logits). This poses a serious threat to LLM-empowered RecSys, which explicitly incorporates sensitive user information, including personal preferences, interaction histories, and user profiles, into the prompts \cite{zhao2024recommender}. 
Output-based inversion is generally less appropriate for attacking LLM-empowered RecSys, as its effectiveness hinges on the semantic richness of the model’s output \cite{zhang2024extracting}. However, recommendation tasks such as \textit{binary classification-based recommendation} typically yield short, semantically sparse responses (e.g., “Yes” or “No”), which substantially limits the applicability of output-based inversion. In contrast, logits-based inversion remains effective across various generative tasks \cite{morris2023language}, raising concerns about its feasibility in compromising RecSys privacy.

Figure~\ref{fig:intro} illustrates the process of LLM-empowered RecSys. Users interact with the RecSys by submitting a generic prompt (e.g., “Recommend me a movie.”). Behind the scenes, an orchestrator retrieves user information and rating history, enriches the input to construct a complete prompt. To protect user privacy, the system encodes the sensitive prompt into embedding representations through a local encoder and transfers them to the LLM for inference, rather than sending the raw prompt.
Since the complete prompt resides solely on the server side of the system, which is usually well protected by various internal protocols, it is hard for external attackers to retrieve it directly. However, together with the final output, the logits generated from the complete prompt are sent to the RecSys for post-processing. Adversaries can easily intercept them via network attacks and reconstruct the complete prompt and thereby expose user information. 
Moreover, this threat extends beyond external attackers; malicious users themselves could access logits associated with their queries and recover the underlying prompts, potentially revealing proprietary business insights that the RecSys provider intends to keep confidential, such as prompt construction strategies and internal logic for processing user information.

Given these concerns, the research community currently lacks a comprehensive understanding of the extent to which LLM-empowe-red RecSys are vulnerable to inversion attacks. To fill this gap, we propose an optimized inversion framework that reconstructs user prompts from LLM-empowered recommendation logits by leveraging a \texttt{vec2text}\cite{morris2023text} engine combined with a similarity-guided refinement procedure. 
{Our optimization method employs beam search to generate candidate prompts and iteratively selects the one that its logits have the highest cosine similarity to the ground truth until convergence. 
This mechanism effectively enhances attack performance and highlights the significant privacy risks posed by inversion attacks in LLM-empowered RecSys.}
Our key contributions in this work are as follows:
\begin{itemize}[leftmargin=*]
    \item We systematically investigate a novel problem: whether existing LLM-empowered RecSys are resilient to inversion attacks, which enabling adversaries to reconstruct detailed prompts from system's output. To the best of our knowledge, this is the first work to investigate the privacy vulnerability of the LLM-empowered RecSys.
    
    \item We propose a synthetic dataset construction pipeline that generates datasets incorporating diverse prompt templates spanning various recommendation tasks. Using this pipeline, we construct two comprehensive, large-scale datasets in the movie and book domains. Leveraging these datasets, our inversion performance improves substantially: at best, the model is able to recover 60\% of user-interacted item titles and infer 85\% of users’ age and gender attributes. The datasets are publicly available at \url{https://github.com/xuemingxxx/Attack_RecSys/}.
    
    \item We introduce a refined method that integrates the \texttt{vec2text} approach with similarity-guided beam search strategy, resulting in an additional 5–13\% improvement in reconstruction fidelity.

    \item Through comprehensive experiments, we further validate the robustness and generality of our approach on two representative LLM-empowered RecSys in the Movie and Book recommendations domain.       
\end{itemize}
\section{PRELIMINARIES}

\begin{figure*}[htbp]
    \centering
    \includegraphics[width=0.88\linewidth]{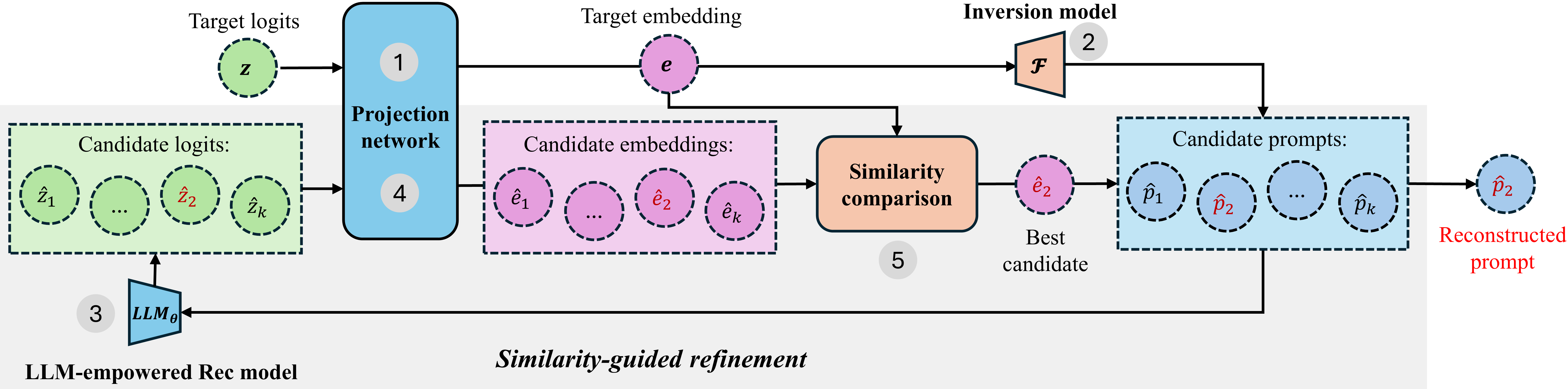}
    \caption[]{Overview of the proposed inversion framework. Given the logits of the target prompt $\mathbf{z}$, our attack contains 5 steps: 1) we first project $\mathbf{z}$ into an embedding $\mathbf{e}$, and then 2) reverse $\mathbf{e}$ into a set of prompt candidates $\{ \hat{p}_1, \dots, \hat{p}_k \}$ with our inversion model $\mathcal{F}$. 3) Those candidates will be input into the LLM-empowered recommendation model and 4) the projection network to get their corresponding embeddings. 5) We finally compare the similarity between each candidate embedding with the target one by cosine similarity. The candidate prompt whose corresponding embedding is closest to the target one will be the final reconstructed prompt.}
    \label{fig:method}
\end{figure*}

In this section, we first introduce the LLM-empowered recommendation model. 
Then, we will define a model inversion attack against LLM-empowered RecSys. 

\subsection{LLM-Empowered Recommendation}
Given a set of users $\mathcal{U} = \{u_1, u_2, \ldots, u_{|\mathcal{U}|}\}$ and a set of items $\mathcal{I} = \{i_1, i_2, \ldots, i_{|\mathcal{I}|}\}$, where each user $u$ has individual profile information $u^{pro}$ like gender and age, and each item $i$ has corresponding textual information $i^{text}$, such as title, brand, and description. 
The goal of a RecSys is to capture users' preferences by modeling the interactions between users and items, e.g., ratings, clicks, and purchases.
Within the framework of a general LLM-empowered RecSys, given an input-output sequence pair $(P, Y)$, a large language model, such as Llama \cite{touvron2023llama}, Qwen \cite{bai2023qwen}, or T5 \cite{raffel2020exploring} serves as the core engine for inferring recommendations from the input $P$. 
This model is denoted as $LLM_\theta$, where $\theta$ represents the model parameters. 
Additionally, $P$ is typically organized by task instruction $P_{task}$, context $P_{context}$, user profile $P_{i}$, and the user’s historical interactions towards items $V^{u_i}=[v_i,v_2,..,v_{}]$. 
Accordingly, a typical input $P$ can be denoted as:
\begin{equation}
    P = [P_{task}; P_{context}; P_{i}; V^{u_i}]
\end{equation}

After that, LLM-empowered RecSys will generate recommendations $Y$ based on the textual input $P$. 
The auto-regressive language generation loss (i.e., Negative Log-Likelihood) is employed to evaluate the discrepancy between the predictions and the target output, defined as follows:
\begin{equation}
\mathcal{L}_{rec}(P,Y) = \frac{1}{|Y|}\sum_{t=1}^{|Y|}-\log p(Y_t|P,Y_{<t}),
\end{equation}
where $\log p(Y_t|P,Y_{<t})$ represents the probability assigned to the item that users are interested in.
$\mathcal{L}_{rec}(P,Y)$ indicates whether the RecSys can predict the target $Y$ and vice versa.

\subsection{Attack Objective}
\label{sec:att_obj}

In this work, we focus on attacking black-box LLM-empowered RecSys, where the internal details of the target system, such as model architecture, gradients, and parameters, are inaccessible.
We assume the adversary has access to the victim's RecSys query interface.  
Specifically, the adversary can raise a large number of queries using fabricated user inputs and obtain the corresponding output logits. 
Moreover, we assume the adversary can obtain some general rating records along with the associated user ID, textual item title and rating score, from publicly available datasets.

\noindent\textbf{Inversion Attack.}
The primary objective of the inversion attack is to reconstruct the original input $P$ to the victim recommender system using only its output, i.e., \begin{equation} InversionAttack(LLM_\theta, \hat{Y}) \xrightarrow{} \hat{P} \xrightarrow{} [\hat{P}_{task}; \hat{P}_{context}; \hat{P}_{i}; \hat{V}^{u_i}]. \end{equation} 

The attack takes the blackbox LLM-based recommendation model $LLM_\theta$ and the target output logits $\hat{Y}$ as input, and aims to infer an approximation $\hat{P}$ of the original input.
Through this process, the adversary can recover sensitive components of the input, including system instruction $P_{task}$, contextual prompts $P_{context}$, the target user's personal information $P_i$, and the interaction history $V^{u_i}$.

\section{METHOD}
\label{sec:method}

\subsection{Overview}
The adversary's goal is to accurately reconstruct the prompt submitted to the LLM-empowered recommendation model. As illustrated in Figure \ref{fig:method}, our proposed inverter framework takes the target logits (i.e., the next-token probability distributions) as the input and processes them in 5 steps with four components: projection network, inversion model, the victim model, and similarity comparison. 

The inversion model and similarity comparison are the main components of our attack. In particular, the inversion model utilizes a \texttt{vec2text} \cite{morris2023language, morris2023text} encoder-decoder model, which maps the logits back to a set of possible textual prompts. However, it requires fixed-size embedding sequences that match its vocab size and input dimensionality, rather than the raw logits produced by the victim model. Thus, in the first step, following the approach of Morris et al. \cite{morris2023language, morris2023text}, we employ a projection network to convert the logits into fixed-size embeddings. 

Rather than outputting a set of possible prompts, we aim to get the most possible one. Thus, our framework also contains a similarity-guided refinement stage. Specifically, we input the candidate prompts into the victim model and the projection network and get their corresponding embeddings, and then we compare each candidate embedding with the target one using cosine similarity. Finally, only the prompt whose embedding is closest to the target one is the final output.

\subsection{Projection and Inversion}
\label{sec:base_inversion}
Let $P \in V^L$ denote the true prompt submitted to the LLM-empowered RecSys, where $V$ is the vocabulary and $L$ is the prompt length. The recommender maps the input prompt into a probability distribution over its vocabulary. In practice, the system produces a logit matrix 
$
\mathbf{z} \in \mathbb{R}^{N \times V},
$
where $N$ is the number of tokens and $V$ is the vocabulary size. After optional top-$k$ and top-$p$ filtering and temperature scaling, 
a projection network aligns the logits with the inversion model's vocabulary indices and padded as necessary to yield a fixed-dimension matrix 
$
\mathbf{h} \in \mathbb{R}^{B \times V'},
$
where $B$ is the batch size.
It then maps $\mathbf{h}$ into a sequence of embedding vectors, resulting in a tensor 
$
\mathbf{e} \in \mathbb{R}^{B \times T \times d},
$
where $T$ is the predetermined sequence length (e.g., 64) and $d$ is the hidden dimensionality. In our framework, this tensor $\mathbf{e}$ represents a processed version of the original logits and is treated as the embedding of the prompt. 

A pretrained encoder–decoder model $\mathcal{F}(\cdot)$ (e.g., a T5-like architecture) then uses these embeddings to generate a textual sequence as the reconstructed prompt 
$
\hat{P} = \mathcal{F}(e).
$ 
This base inversion confirms that the recommender’s logits, although intended for next-token prediction, retain rich semantic details sufficient to recover large portions of the original prompt.

\subsection{Similarity-Guided Refinement}
To further enhance the accuracy of prompt reconstruction, we employ a \emph{similarity-guided refinement} strategy that iteratively optimizes the candidate prompt generated by the inversion model. In this stage, the objective is to select the candidate prompt that best aligns with the target embedding \( e \).

At each iteration \( t \), the inversion model generates a set of \( K \) candidate prompts via beam search:
\[
\left\{\hat{P}_1^{(t)},\,\hat{P}_2^{(t)},\,\ldots,\,\hat{P}_K^{(t)}\right\},
\]
where each candidate \( \hat{P}_k^{(t)} \) is generated by the inversion model and \( K \) denotes the beam width. 
To assess the semantic consistency between each candidate and the target prompt, we calculate the cosine similarity between the candidate embedding, obtained via the victim recommendation model \( LLM_\theta(\cdot) \), and the target embedding \( e \). Specifically, the candidate that maximizes the cosine similarity is selected as the updated hypothesis:

\begin{equation}
\hat{P}^{(t+1)} = \arg\max_{1 \leq k \leq K} \cos\Bigl(LLM_\theta\bigl(\hat{P}_k^{(t)}\bigr),\,e\Bigr),
\end{equation}
where \(LLM_\theta(\hat{P}_k^{(t)})\) represents the logits produced by the rec model for the candidate prompt \( \hat{P}_k^{(t)} \). This update rule ensures that the hypothesis is progressively refined to be closer to \( e \) in the embedding space.

The iterative process is terminated when the improvement in cosine similarity between two consecutive hypotheses falls below a predefined threshold \( \epsilon \):

\begin{equation}
\cos\Bigl(LLM_\theta\bigl(\hat{P}^{(t)}\bigr),\,e\Bigr) - \cos\Bigl(LLM_\theta\bigl(\hat{P}^{(t-1)}\bigr),\,e\Bigr) < \epsilon.
\end{equation}

In practice, we set \( \epsilon = 1 \times 10^{-5} \) to ensure convergence. The final refined prompt, denoted by \( \hat{P}^{(T)} \) after \( T \) iterations or upon satisfying the stopping criterion, is regarded as the reconstructed approximation of the recommender's original input prompt.

Empirically, we observe that setting the beam width $k = 5$ yields the best performance. In most cases, the cosine similarity between candidate and target logits converges within two iterations, indicating that a single round of beam search is often sufficient to achieve near-optimal reconstruction. Nevertheless, the iterative refinement mechanism serves as an additional safeguard, enhancing reconstruction accuracy when dealing with more complex or nuanced queries.

\section{SYNTHETIC DATASET CONSTRUCTION}
\label{sec:synthetic_method}

The effectiveness of an inversion attack depends on training an effective inversion model, which relies heavily on appropriate data with prompts and corresponding logits. However, there is no such dataset in the recommendation domain. Moreover, the model trained on general-purpose LLM datasets, such as the \emph{one-million-instructions} corpus \cite{morris2023language}, proves inadequate for attacking RecSys, as later discussed in Section \ref{sec:results}. Given the lack of a large and domain-specific benchmark for training effective inversion attack models for LLM-empowered RecSys, and the fact that adversaries in black-box scenarios have no knowledge to the victim model’s exact prompt templates, we design a synthetic dataset construction pipeline. This pipeline aims to generate datasets that incorporate diverse prompt templates across multiple mainstream LLM-based recommendation tasks so that the inversion models trained on such datasets can effectively attack a wide range of LLM recommenders specialized in different types of recommendation tasks and prompt formats. 
Notably, the logits for each prompt are generated dynamically by the respective victim recommender and are thus excluded from the dataset construction pipeline.

As outlined in Section~\ref{sec:att_obj}, we mentioned that the adversary has access to publicly available datasets on other platforms containing rating records with the essential attributes: \texttt{userId}, \texttt{itemId}, \texttt{item\_title}, and \texttt{rating}. To ensure the resulting dataset is applicable to a broad range of LLM-empowered RecSys, we collect instruction templates covering five representative recommendation tasks: \textit{binary classification-based recommendation}, \textit{direct recommendation}, \textit{sequential recommendation}, \textit{rating prediction}, and \textit{cold start recommendation} \cite{wang2024towards,zhao2024recommender}, with more than ten templates for each task, we randomly sample an instruction template from the corresponding task-specific pool during dataset construction.

We first preprocess the public data by grouping records by \texttt{userId}, collecting each user's rated item titles along with their corresponding ratings, and sorting them in reverse chronological order based on timestamp. For recommendation tasks that require distinguishing between user preferences and non-preferences, we apply a threshold parameter $k$: items with ratings greater than or equal to $k$ are treated as preferred, whereas those with ratings below $k$ are considered non-preferred. To control the complexity of input prompts, we introduce a parameter $n$ that specifies the maximum number of item titles included in each instruction. For tasks involving user profile information, if demographic attributes such as age and gender are available in the dataset, we embed them into the prompts using fixed template phrases, e.g., \textit{``The user is a <age>-year-old <gender>.''} When such attributes are unavailable, synthetic demographic values are randomly assigned to ensure diversity and completeness in the prompt construction.
The complete data-generation workflow is summarized in Algorithm~\ref{alg:gen_dataset}.

\begin{algorithm}[ht]
\caption{Instruction Dataset Construction for Inversion Attack}
\label{alg:gen_dataset}
\KwIn{Public dataset $\mathcal{D}$ with fields \texttt{user\_Id}, \texttt{item\_Id}, \texttt{item\_title}, \texttt{rating}, optionally \texttt{age}, \texttt{gender}; rating threshold $k$; item limit per prompt $n$; instruction templates $\mathcal{T}$ for each task type}
\KwOut{Constructed instruction set $\mathcal{S}$}
$\mathcal{S} \gets \emptyset$ \tcp*[r]{Initialize the instruction set}

Group $\mathcal{D}$ by \texttt{user\_Id} to obtain per-user subsets $\{\mathcal{D}_u\}$\;

\ForEach{user $u$ with records $\mathcal{D}_u$}{
    Sort $\mathcal{D}_u$ in descending order of timestamp\tcp*[r]{Sort user interactions by recency}
    
    \If{demographics available in $\mathcal{D}_u$}{
        $age_u \gets$ recorded age\;
        $gender_u \gets$ recorded gender\;
    }
    
    Sample $age_u \sim \mathcal{U}(18, 65)$\tcp*[r]{Random age if not recorded}
    Sample $gender_u \sim \{\text{male}, \text{female}\}$\tcp*[r]{Random gender if not recorded}

    $\mathcal{I}_u^+ \gets \{i \in \mathcal{D}_u \mid rating_i > k \}$\tcp*[r]{High-rated items}
    $\mathcal{I}_u^- \gets \{i \in \mathcal{D}_u \mid rating_i < k \}$\tcp*[r]{Low-rated items}

    Sample up to $n$ items from each of $\mathcal{I}_u^+$ and $\mathcal{I}_u^-$\;

    \ForEach{task type $t \in \{\text{Binary, Direct, Sequential, Rating, Cold-start}\}$}{
        Select template $T$ from $\mathcal{T}_t$ uniformly at random\tcp*[r]{Randomly choose a template for this task type}
        
        Fill $T$ with user profile $(age_u, gender_u)$ and the sampled item sets\;
        
        $s \gets \textsc{Format}(T, u)$\tcp*[r]{Convert template into formatted instruction}
        $\mathcal{S} \gets \mathcal{S} \cup \{s\}$\;
    }
}
\Return $\mathcal{S}$\;
\end{algorithm}
\section{EXPERIMENTS}
Our study aims to evaluate the vulnerability of LLM-empowered RecSys to inversion attacks through comprehensive experiments. 
Specifically, we investigate the following research questions:
    \textbf{RQ1}: Are existing LLM-empowered RecSys resilient to inversion attacks?
    \textbf{RQ2}: Are the attacks generalizable across domains and different types of LLM-empowered RecSys?
    \textbf{RQ3}: What are the key factors contributing to a successful inversion attack?
    \textbf{RQ4}: How does the recommendation performance of RecSys influence the effectiveness of inversion attacks?
    \textbf{RQ5}: What are the limitations of the attack method?

\subsection{Experimental Setup}
\subsubsection{Victim Recommendation Models}
\label{sec:victim_models}
We utilize two popular LLM-based recommender models with distinct architectures, \textbf{TallRec} \cite{bao2023tallrec} and \textbf{CoLLM} \cite{zhang2025collm}, to investigate the privacy vulnerability of LLM-empowered RecSys:

\begin{itemize}[leftmargin=*]
  \item \textbf{TallRec} \cite{bao2023tallrec}. It is an \emph{ID-free} recommendation model that uses titles to represent items and convert the user-item interactions to a language format. It fine-tunes a frozen LLaMA-7B model using LoRA, framing recommendation as a binary classification task: given a textual prompt encoding a user's interaction history, the model predicts whether the user will like a target item.
  
  \item \textbf{CoLLM} \cite{zhang2025collm}. It enhances LoRA-tuned for recommendation by integrating collaborative signals from user-item interactions. It encodes user and item embeddings from traditional collaborative models and maps them into the LLM’s token space, allowing the model to leverage both textual and collaborative information. In our experiments, we follow the original setup and adopt the Qwen-7B model as the backbone to train CoLLM-based recommendation models.
  
\end{itemize}

\subsubsection{Datasets for RecSys}
\label{sec:data_recsys}
Two real-world datasets are used to construct two recommendation scenarios.

 \begin{itemize}[leftmargin=*]
     \item \textbf{Movie Scenario.} We employ one widely used movie dataset \textsc{ML-1M}\footnote{\url{https://grouplens.org/datasets/movielens/1m/}} from the \textsc{MovieLens} collection \cite{harper2015movielens}.

     The data is partitioned into a training set (80\%), a validation set (10\%), and a test set (10\%).  We train recommendation models using the \textbf{TallRec} and \textbf{CoLLM} method on the training and validation sets, thereby obtaining two distinct movie LLM-based RecSys. 

    \item \textbf{Book Scenario.}
    For the book domain, we adapt the \textsc{Amazon Books Reviews}\footnote{\url{https://www.kaggle.com/datasets/mohamedbakhet/amazon-books-reviews/}} dataset from Kaggle, which comprises 3 million ratings across 212 thousand book titles. Following the same experimental protocol in the movie scenario, we train and evaluate book recommendation models with both \textbf{TallRec} and \textbf{CoLLM}, producing two distinct LLM-based RecSys for the book domain. 

\textbf{Notably}, the training and validation data on these two datasets are used for the target victim recommendation model training only and are NOT visible to the inversion model. The test set on these two datasets are used for evaluating attack effectiveness, simulating a realistic adversarial scenario.

\end{itemize}

\subsubsection{Datasets for Inversion Model.}
\label{sec:data_inversion}
To train a robust inversion attack model, it is essential to synthesize a large-scale corpus tailored to the target domain. For the movie and book recommendation domains, we leverage the other publicly available datasets (not the datasets used for recommendation model training) as source data to construct the training corpus. The details are described as follows.

 \begin{itemize}[leftmargin=*]
 
     \item \textbf{Movie Scenario.}
     We employ two other movie datasets  \textsc{ML-25M}\footnote{\url{https://grouplens.org/datasets/movielens/25m/}} and \textsc{ML-32M}\footnote{\url{https://grouplens.org/datasets/movielens/32m/}} from the \textsc{MovieLens} collection as source data. Following the generation procedure outlined in the Algorithm~\ref{alg:gen_dataset}, as the rating scores in both datasets are on a scale of 0 to 5, we set the threshold $k$ to 4 for distinguishing preferred and non-preferred items. To enhance the robustness and realism of the constructed dataset, the number of item titles included in each prompt $n$ is randomly sampled from the interval $(3, 11)$.
 This process yields a large-scale corpus comprising 1.5 million prompts, which we refer to as the $InvInst_{movie}$ dataset.
 
     \item \textbf{Book Scenario.} 
     We utilize the \textsc{1M Users Book Ratings}\footnote{\url{https://www.kaggle.com/datasets/paramamithra/1m-users-book-ratings/data}} and \textsc{Goodreads Book Reviews}\footnote{\url{https://www.kaggle.com/datasets/pypiahmad/goodreads-book-reviews1/}}  datasets from the Kaggle platform as source data. Applying the same generation strategy, we construct a corresponding corpus of 1.5 million prompts, denoted as the $InvInst_{book}$ dataset.
     
\end{itemize}

\begin{table*}[!htbp]
\centering
\setlength{\tabcolsep}{2pt}
\renewcommand{\arraystretch}{1.1}
\caption[]{Main results of attacking \textbf{TallRec} on Movie and Book domains. The best results are highlighted in bold.}
\label{tab:major_results_tall}
\resizebox{0.95\textwidth}{!}{%
\begin{tabular}{lccccccccccc}
\toprule
\textbf{Models} & \multicolumn{5}{c}{\textbf{Movie}} & & \multicolumn{5}{c}{\textbf{Book}} \\
\cmidrule{2-6} \cmidrule{8-12}
& ItemMatch & ProfileMatch & BLEU & ROUGE & Token-level F1 & & ItemMatch & ProfileMatch & BLEU & ROUGE & Token-level F1 \\
\midrule
\texttt{Generic} & 0.0011 & 0 & 1.5196 & 0.1436 & 0.1986 & & 0.0015 & 0 & 1.4305 & 0.1580 & 0.1919 \\
\texttt{Generic-Rec} & 0 & 0 & 1.6056 & 0.1753 & 0.2267 & & 0.0004 & 0 & 1.5266 & 0.1975 & 0.2311 \\
\texttt{RecAtk-Spec(Ours)} & 0.6086 & 0.8550 & 84.89 & 0.9150 & 0.9056 & & 0.3277 & 0.8337 & 70.3573 & 0.7803 & 0.7665 \\
\texttt{RecAtk-Refined(Ours)} & \textbf{0.6471} & \textbf{0.87} & \textbf{85.8427} & \textbf{0.9236} & \textbf{0.9175} & & \textbf{0.3463} & \textbf{0.8337} & \textbf{70.4409} & \textbf{0.7818} & \textbf{0.7678} \\
\bottomrule
\end{tabular}
}
\end{table*}

\begin{table*}[!htbp]
\centering
\setlength{\tabcolsep}{2pt}
\renewcommand{\arraystretch}{1.1}
\caption[]{Main results of attacking \textbf{CoLLM} on Movie and Book domains. The best results are highlighted in bold.}
\label{tab:major_results_co}
\resizebox{0.95\textwidth}{!}{%
\begin{tabular}{lccccccccccc}
\toprule
\textbf{Models} & \multicolumn{5}{c}{\textbf{Movie}} & & \multicolumn{5}{c}{\textbf{Book}} \\
\cmidrule{2-6} \cmidrule{8-12}
& ItemMatch & ProfileMatch & BLEU & ROUGE & Token-level F1 & & ItemMatch & ProfileMatch & BLEU & ROUGE & Token-level F1 \\
\midrule
\texttt{Generic} & 0.0014 & 0 & 2.1154 & 0.1362 & 0.1873 & & 0.0036 & 0 & 2.5638 & 0.1598 & 0.1959 \\
\texttt{Generic-Rec} & 0.0001 & 0 & 1.3563 & 0.1534 & 0.2026 & & 0.0002 & 0 & 1.4434 & 0.1771 & 0.2122 \\
\texttt{RecAtk-Spec(Ours)} & 0.4741 & 0.4463 & 66.3841 & 0.7598 & 0.7660 & & 0.2297 & 0.2110 & 51.0628 & 0.6265 & 0.6515 \\
\texttt{RecAtk-Refined(Ours)} & \textbf{0.5125} & \textbf{0.5012} & \textbf{67.5750} & \textbf{0.7698} & \textbf{0.7760} & & \textbf{0.2612} & \textbf{0.2830} & \textbf{53.1562} & \textbf{0.6468} & \textbf{0.6628} \\
\bottomrule
\end{tabular}
            }
\end{table*}

\subsubsection{Evaluation Metrics}
\label{sec:metrics}

To evaluate the robustness of LLM-empo-wered RecSys against the proposed inversion attack, we adopt three metrics, BLEU, ROUGE-L, and Token-level F1 from the work \cite{morris2023language} to measure similarity between generated and reference texts.
Additionally, we introduce two novel metrics tailored to recommendation scenarios.

\begin{itemize}[leftmargin=*]
    
    \item ItemMatch.
    It reflects the proportion of ground-truth items that are successfully recovered in the reconstructed prompt, irrespective of their order, defined as: 
    \begin{equation}
    \operatorname{ItemMatch} \;=\; 
    \frac{\lvert \mathcal{T} \cap \hat{\mathcal{T}} \rvert}{\lvert \mathcal{T} \rvert}\times 100\%,
    \end{equation}
    where $\mathcal{T}$ and $\hat{\mathcal{T}}$ denote the sets of item titles in the reference and reconstructed prompts, respectively.
    
    \item ProfileMatch.
    It evaluates the leakage of sensitive user information. Formally, let $N_{\text{total}}$ denote the number of prompts that contain demographic attributes, and $N_{\text{correct}}$ represent the subset of cases in which both age and gender are recovered exactly.
    \begin{equation}
    \operatorname{ProfileMatch} \;=\;
    \frac{N_{\text{correct}}}{N_{\text{total}}}\times 100\%.
    \end{equation}
\end{itemize}

\subsubsection{Baselines and Our Methods}
\label{sec:baselines}

In the absence of prior work on inversion attack for LLM-based RecSys, we establish baselines by adapting the public \texttt{vec2text} framework \cite{morris2023language,morris2023text}. All inversion models below share the same T5-base \cite{raffel2020exploring} architecture as their backbone.

\begin{itemize}[leftmargin=*]
  \item \textbf{\texttt{Generic}}. 
        The original inversion model trained to reconstruct prompts from \textsc{LLaMA-7B} \cite{morris2023language}, applied directly to attack RecSys without modification. This model was trained on the general \emph{one-million-instructions} corpus\footnote{\url{https://huggingface.co/datasets/wentingzhao/one-million-instructions}}.

  \item \textbf{\texttt{Generic-Rec}}.  
        A fine-tuned version of the \texttt{Generic} model where we maintain the same training corpus but replace the logit source with our victim LLM-powered RecSys instead of vanilla LLaMA-7B.
          
  \item \textbf{\texttt{RecAtk-Spec (Ours)}}.  
        Our specialized approach fine-tunes the inversion model using domain-adapted datasets $InvIst_{movie}$ or $InvIst_{book}$. 

  \item \textbf{\texttt{RecAtk-Refined (Ours)}}.
        An enhanced version of \texttt{RecAtk-Spec} incorporating the Similarity-Guided Refinement mechanism for improved attack performance.
  
\end{itemize}

\subsubsection{Implementation Settings}
\label{sec:implementation}
We utilize T5-base \cite{raffel2020exploring} as the backbone model of all inversion models, which has 222M parameters. The maximum length of the input sequence is set to 256 tokens.
Models are trained with a learning rate of $2\mathrm{e}^{-4}$ using a constant schedule and a linear warmup over the first 100,000 steps.
All models are trained using \texttt{bfloat16} precision.
Training is conducted for up to 50 epochs with early stopping based on validation performance. Specifically, ItemMatch on the validation set is evaluated every 10,000 steps, and training is terminated if its improvement is less than $10^{-3}$ for five consecutive evaluations.

\subsection{Attack Effectiveness (RQ1 \& RQ2)}
\label{sec:results}

We evaluate the attack performance against both \textbf{TallRec} and \textbf{CoLLM} using baseline models \texttt{Generic} and \texttt{Generic-Rec}, as well as our proposed optimization models \texttt{RecAtk-Spec} and \texttt{RecAtk\-Refined}. 
The results are presented in Table \ref{tab:major_results_tall} and Table \ref{tab:major_results_co}, respectively. From these results, we observe the following findings:

\begin{itemize}[leftmargin=*]
    \item 
    \textbf{Strong Prompt Reconstruction Ability.}
Our models, \texttt{RecAtk\-Spec} and \texttt{RecAtk-Refined}, exhibit a strong ability to reconstruct prompts across both Movie and Book domains. All text similarity metrics, BLEU, ROUGE, and Token-level F1, exceed $51\%$, with values surpassing $85\%$ on the TallRec-based Movie recommendation model. 
Moreover, \texttt{RecAtk-Refined} achieves ItemMatch scores of up to $64\%$ at the best-case scenario, indicating its ability to recover meaningful user interaction histories or personal preferences. These results reveal the susceptibility of current LLM-based RecSys to prompt inversion attacks.

\item \textbf{High-fidelity Recovery of Sensitive Profiles.}
 \texttt{RecAtk-Spec} and \texttt{RecAtk-Refined} recovery user profile information with striking precision, pushing ProfileMatch to 83.3\%--87.0\%. This improvement indicates that demographic attributes are the most consistently leaked components, likely due to their brevity, constrained vocabulary, and fixed positional structure within the prompt. Such low-diversity fields generate strong and easily learnable signals in the RecSys’s output logits, rendering them highly susceptible to accurate reconstruction.

\item \textbf{Obvious Performance Gap Between Domains.}
We observe a pronounced domain discrepancy both in Table~\ref{tab:major_results_tall} and Table~\ref{tab:major_results_co}: across all metrics on movies is substantially higher than on books. The ItemMatch rate reaches 65\% for movie titles, whereas only 35\% for books, and similar declines are observed in BLEU, ROUGE, and token-level F1 scores. To investigate the underlying causes, we compiled corpus-level statistics from both the training and test sets. Among the 2,265 unique movie titles in the test set, 96\% (2,178) are also present in the training corpus. In contrast, only 83\% (4,158 out of 5,038) of book titles overlap, leaving 880 titles entirely unseen during training. In addition, average tokenized title lengths differ significantly between domains: movie titles span an average of 4.87 LLaMA tokens, whereas book titles average 8.93 tokens. Long-form titles pose additional challenges for inversion, book titles exceeding 20 tokens are almost never recovered. 

These factors collectively account for the observed performance gap. In the movie domain, the high lexical overlap between train and test sets enables the inversion model to memorize an embedding for almost every target title, yielding high recovery rates. In contrast, the book domain presents a larger share of unseen titles and markedly longer sequences. Because residual information in a model’s next-token distribution attenuates rapidly with both token distance and word frequency, the cues associated with these long or rare book titles are pushed into low-probability tails and become difficult to retrieve \cite{morris2023language}.

\item \textbf{Attack Difficulty Varies across RecSys.}
Our experiments reveal significant variation in attack difficulty across different recommendation systems. While the attack maintains effectiveness across all evaluation metrics, we observe consistently stronger performance against TallRec compared to CoLLM. This discrepancy stems from CoLLM's collaborative filtering layer, which introduces interference that partially distracts the inversion model's attention. Consequently, inversion performance experiences a modest degradation, with ProfileMatch accuracy showing the most pronounced decline. Notably, despite these challenges, our results demonstrate that the inversion method remains both effective and transferable across diverse LLM-based RecSys with different backbone models and structural components.

\end{itemize}

\subsection{Contributing Factors to Attack Success (RQ3)}
As demonstrated in Section~\ref{sec:results}, LLM-empowered RecSys are highly susceptible to inversion attacks. To understand what contributes to the success of such attacks, we further analyze the results from Table~\ref{tab:major_results_tall} and Table~\ref{tab:major_results_co}.

\begin{itemize}[leftmargin=*]

\item \textbf{Domain Alignment is Critical.}
Inversion models trained on generic instruction data (\texttt{Generic}, \texttt{Generic-Rec}) fail to reconstruct prompts (ItemMatch $\approx$ 0\%, ProfileMatch = 0\%). In contrast, models trained on domain-aligned data (\texttt{RecAtk-Spec}, \texttt{RecAtk\-Refined}) achieve substantial gains, e.g., ItemMatch reaches 61\% in the movie domain and 33\% in the book domain; ProfileMatch exceeds 83\%. These findings validate the effectiveness of our domain-specific synthetic dataset.

This performance gap aligns with the residual-information theory by Morris et al.~\cite{morris2023language},  which posits that when test-time prompts contain a substantial proportion of previously unseen content, the mapping between low-order statistics in the next-token probability vector and the lexical “slots” learned during training collapses. As a result, the \texttt{Generic-Rec} inverter which trained on a corpus lacking domain alignment, faces numerous out-of-vocabulary item titles and profile information, which prevents its decoder from reliably projecting residual logits back into the original prompt space. Conversely, fine-tuning on the domain-adapted $InvInst_{movie}$ corpus restores this alignment, elevating reconstruction performance from near-zero to substantial levels.

\item \textbf{Similarity-Guided Refinement Mechanism further enhances performance.}
 \texttt{RecAtk-Refined} consistently outperforms \texttt{RecAtk\-Spec}, yielding a relative improvement of 5–13\% in ItemMatch and approximately 1\% gains in BLEU, ROUGE, and F1 scores in the movie domain, with analogous improvements observed in the book domain. These results demonstrate that Similarity-Guided Refinement enables more precise reconstruction of the original instructions by reframing the generation process from "constructing a plausible sequence token by token" to "locating the point in high-dimensional logit space that is closest to the true distribution." After each round of beam search, instead of blindly accepting the top-probability candidate, we re-query the victim model and evaluate the global consistency between the candidate logits and the original logits using cosine similarity. The candidate with the highest alignment is then selected to achieve the most reliable reconstruction.

\begin{figure}[h]
    \centering
    \begin{subfigure}{0.5\linewidth}
        \centering
        \includegraphics[width=\linewidth]{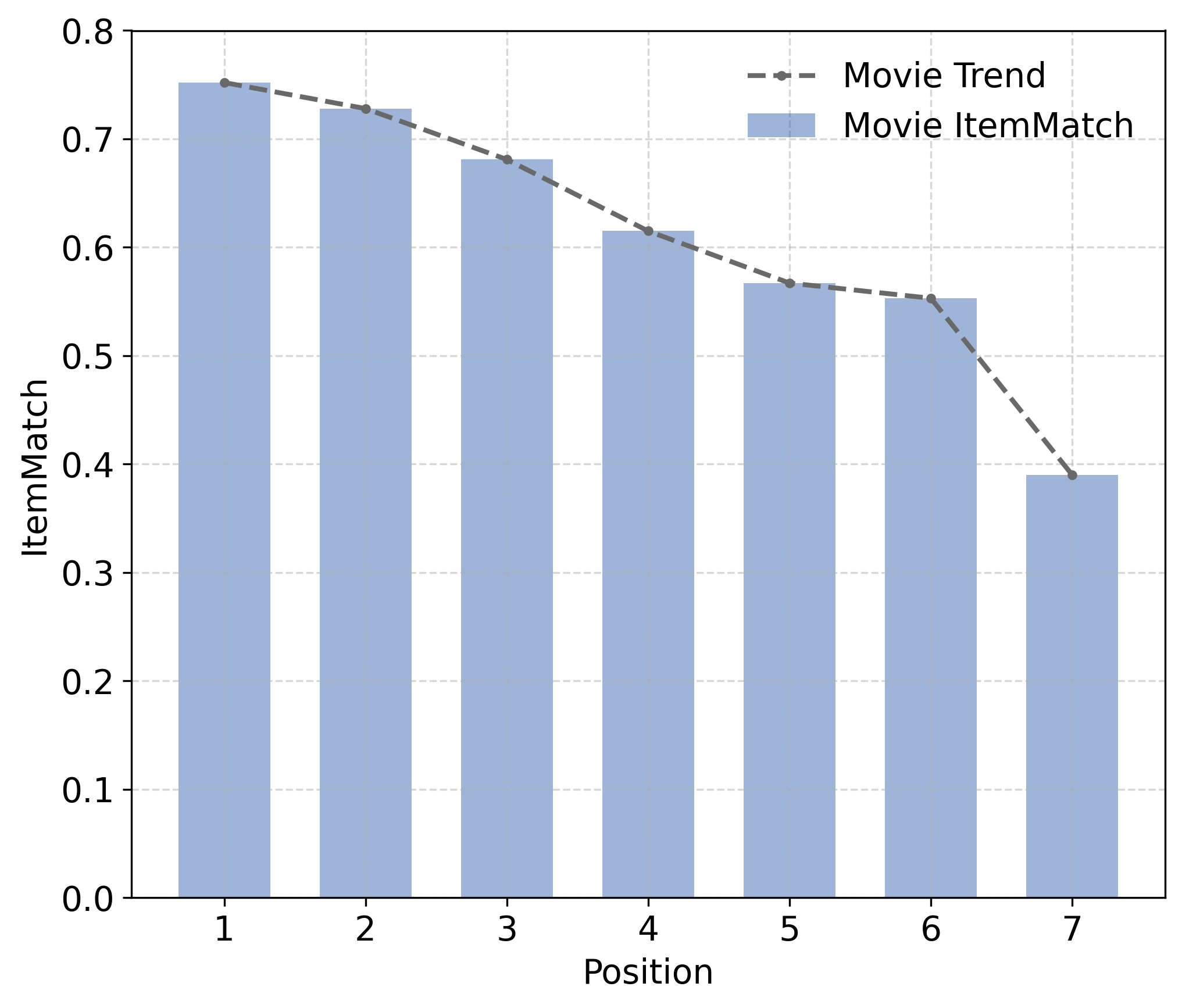}
        \caption{Movie Scenario}
        \label{fig:effect_item_number_movie}
    \end{subfigure}
    \hfill
    \begin{subfigure}{0.49\linewidth}
        \centering
        \includegraphics[width=\linewidth]{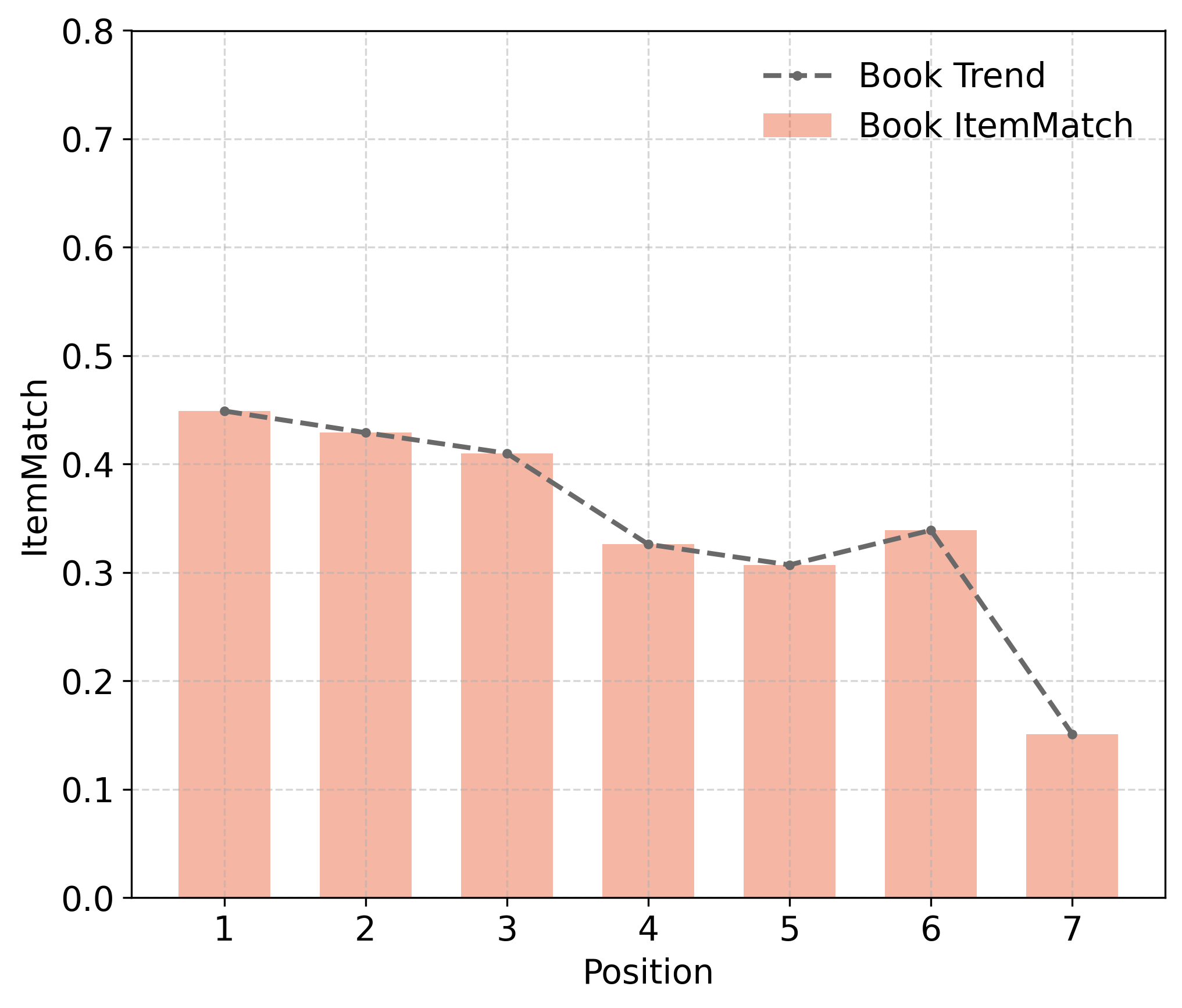}
        \caption{Book Scenario}
        \label{fig:effect_item_number_book}
    \end{subfigure}
    \caption[]{Impact of item position on inversion performance in Movie and Book domains under attacking \textbf{TallRec} scenario. ItemMatch scores are reported for each position, with position 1 indicating the first item in the prompt.}
    \label{fig:distance_effect}
\end{figure}

\item \textbf{Impact of Item Position on Reconstruction Accuracy.}
To further investigate the impact of item position, we conduct additional experiments under the TallRec attack setting. These additional experiments report the recovery success rate for items at each position across the test set. As shown in Figure~\ref{fig:distance_effect}, items appearing earlier in the prompt exhibit higher recovery rates, which gradually decline as the position moves toward the end and a sharp drop is observed at the final position. This pattern does not entirely align with our initial expectations. According to Morris et al.’s study on “Residual Information in LM Logits” \cite{morris2023language}, the farther position is from the last token, the weaker the residual signal it leaves in the next-token probability vector, implying that earlier tokens should be harder to reconstruct. Notably, the sharp drop at the final position aligns with a pattern we observe in the figure of their experiment which plots KL divergence and Hamming distance against the token-wise distance \cite{morris2023language}. The residual signal appears weakest for tokens in positions 4–6 preceding the final token—precisely where our last item resides in the prompt structure. These results suggest that, under our experimental setting, proximity to the logits generation step does not necessarily imply greater ease of inversion.

\end{itemize}

\begin{table}[h]
\centering
\setlength{\tabcolsep}{1pt}
\small
\renewcommand{\arraystretch}{1.1}
\caption{Impact of Victim Recommender Quality on Inversion Performance. The experiment is conducted in the movie domain under the TallRec attack setting.}
\resizebox{0.48\textwidth}{!}{
\begin{tabular}{lccccc}
\toprule
\textbf{Model} & ItemMatch & ProfileMatch & BLEU & ROUGE & Token-level F1 \\
\midrule
\texttt{RecAtk-Refined} & 0.6471 & 0.870 & 85.8427 & 0.9236 & 0.9175 \\
\texttt{RecAtk-Refined-VPD} & 0.6252 & 0.827 & 82.8853 & 0.9174 & 0.9075 \\
\bottomrule
\end{tabular}
}
\label{tab:ablation_study}
\end{table}

\subsection{Relationship Between Attack and Recommendation Performance (RQ4)}
To investigate the correlation between attack effectiveness and recommendation performance, we conduct a controlled experiment comparing our original \textbf{\texttt{RecAtk-Refined}} with a variant when attacking \textbf{TallRec} in the movie domain.

\begin{itemize}[leftmargin=*]
    \item \textbf{\texttt{RecAtk-Refined-VPD}} (Victim Performance Degrading): We compromise TallRec's recommendation quality by contaminating its training data (Section~\ref{sec:data_recsys}) with randomly generated prompts, resulting in a significant AUC drop from 0.6576 to 0.5889. This variant tests how victim model degradation affects attack success.
\end{itemize}

The results in Table~\ref{tab:ablation_study} demonstrate that when the same inversion model is applied to a degraded recommender, the performance remains largely intact: ItemMatch declines only slightly (from 0.647 to 0.625), and ProfileMatch remains above 0.82. BLEU, ROUGE, and F1 each drop by less than two absolute points. These results suggest that \textit{prompt leakage is largely insensitive to the recommender’s performance}, as the logits continue to encode specific details of the input regardless of the overall performance of the RecSys. The slightly noisier logits produced by the weakened recommender still retain token-level signals that the inversion model can exploit effectively.

\subsection{Limitation Analysis (RQ5)}
\begin{figure}[h]
    \centering
    \begin{subfigure}{0.5\linewidth}
        \centering
        \includegraphics[width=\linewidth]{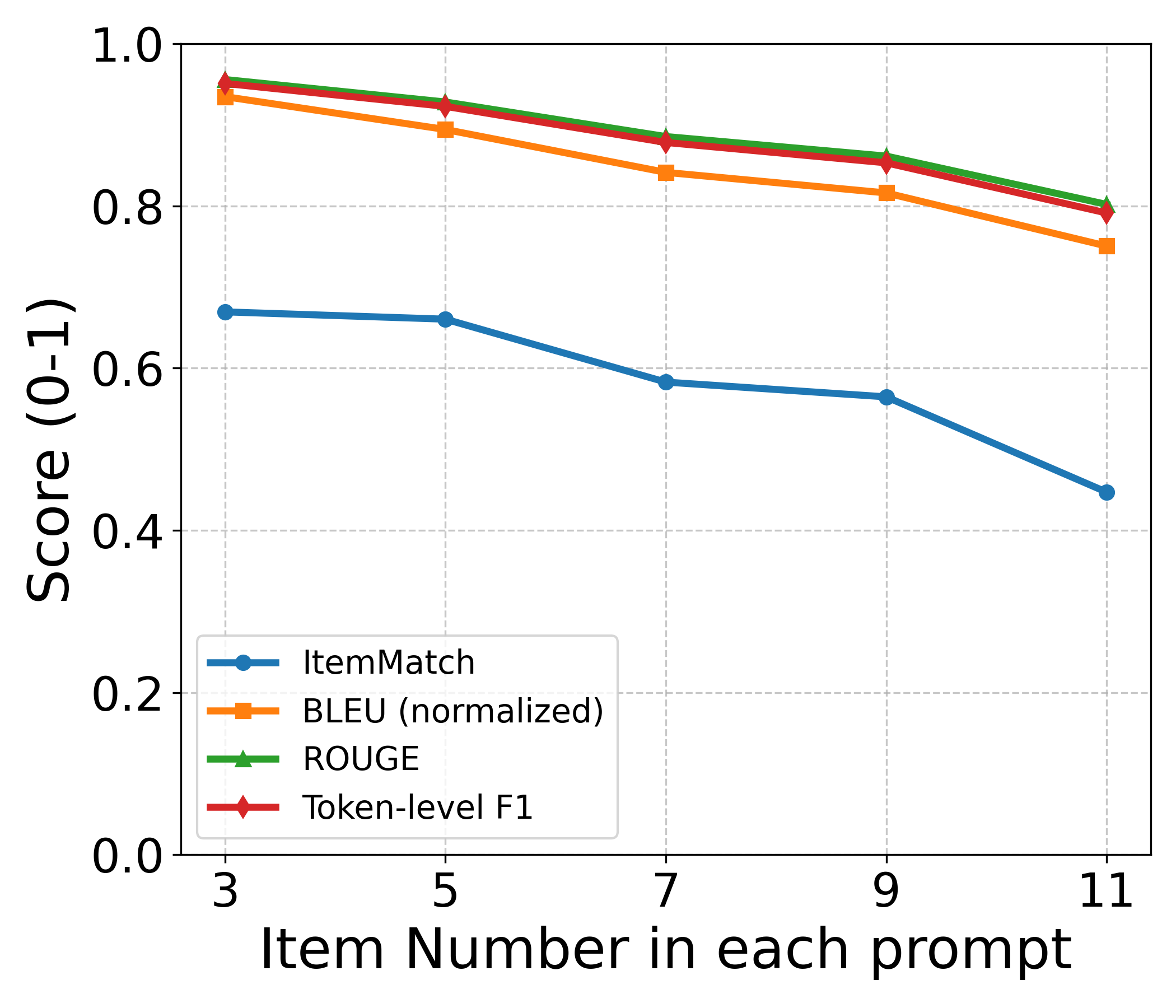}
        \caption{Movie Scenario}
        \label{fig:effect_item_number_movie}
    \end{subfigure}
    \hfill
    \begin{subfigure}{0.49\linewidth}
        \centering
        \includegraphics[width=\linewidth]{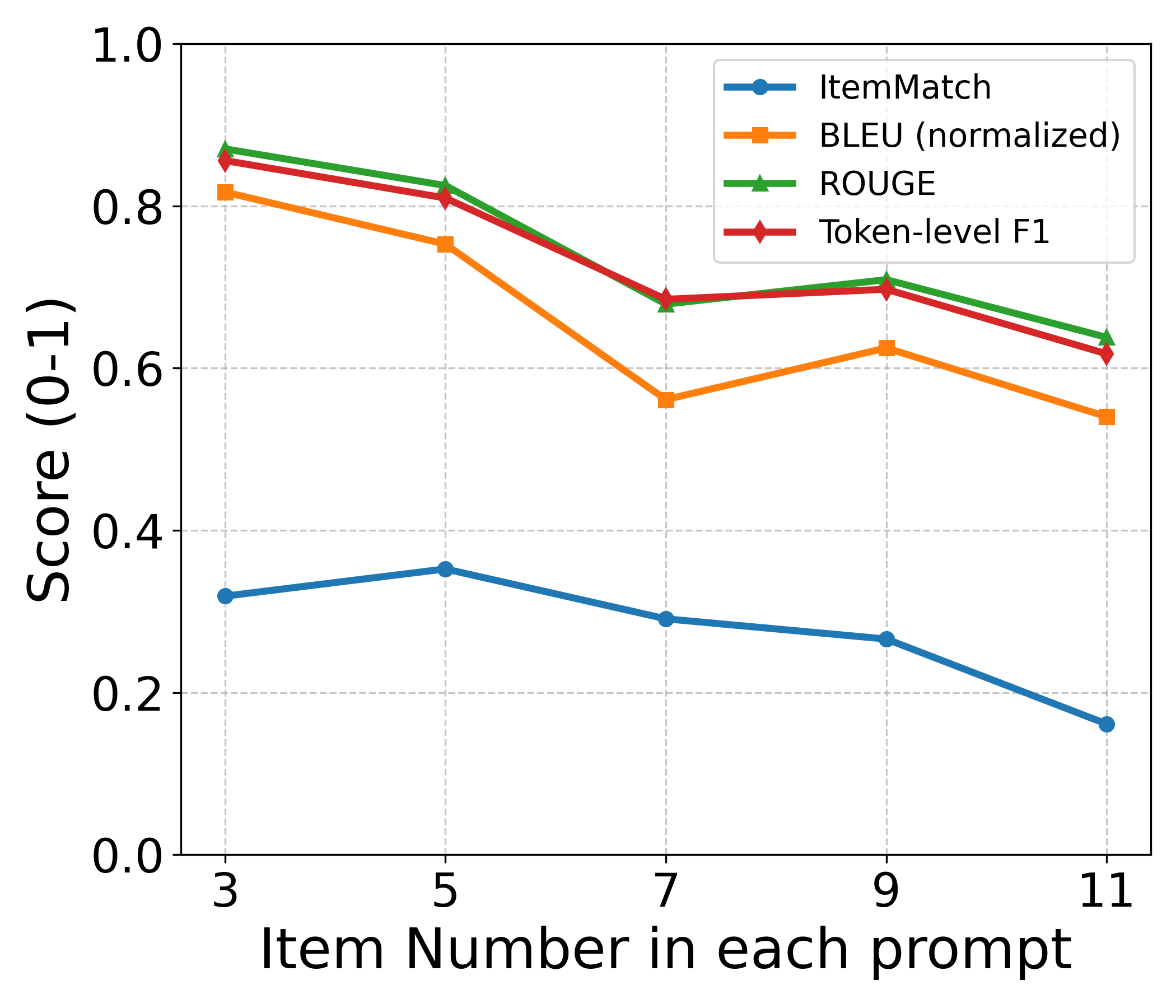}
        \caption{Book Scenario}
        \label{fig:effect_item_number_book}
    \end{subfigure}
    \caption{Effect of item number in prompts on inversion performance across Movie and Book domains (BLEU scores are normalized to 1 for visualization purposes).}
    \label{fig:item_number_effects}
\end{figure}

Despite the overall success of inversion attacks, a key limitation persists: performance deteriorates as prompt length increases. In particular, reconstructing longer prompt sequences poses significant challenges. To examine this effect, we evaluate \texttt{RecAtk-Refined} in the TallRec attack setting on five test sets containing prompts with varying numbers of item titles (3, 5, 7, 9, and 11).
As shown in Figure~\ref{fig:item_number_effects}, inversion fidelity decreases significantly as prompt length increases. All four evaluation metrics decline by approximately 20\% to 24\% as the number of item titles grows from three to eleven. 
This degradation stems from the fact that longer prompts introduce greater semantic variability, yet the inversion model must reconstruct more tokens from a fixed-size probability vector. This constraint dilutes the semantic signal encoded in each dimension, weakening the correspondence between the logits and the original prompt content, and ultimately reducing reconstruction accuracy. These findings suggest that the proposed inversion attack has inherent limitations with respect to prompt length: when prompts become excessively long or contain dense and complex semantic content, model performance tends to degrade accordingly.

\section{RELATED WORK}
In this section, we first review existing LLM-empowered RecSys methods, followed by a discussion of prior work on attacks targeting both LLMs and recommender systems.

\subsection{LLM-Empowered RecSys}
Large language models (LLMs) have recently been adapted to end-to-end recommendation by transforming user–item interactions into natural language prompts and applying lightweight tuning. Xu et al. \cite{xu2024prompting} divide the LLM prompt into four components: task description, user interest modeling, candidate construction, and prompting strategy, and they systematically evaluate how model scale, context length, and tuning methods affect accuracy \cite{xu2024tapping}. Geng et al. \cite{geng2022recommendation} convert all interactions and queries into T5 prompts in P5, achieving strong zero and few-shot performance. Bao et al. \cite{bao2023tallrec} present TALLRec, which fine-tunes LLaMA on recommendation tasks with parameter-efficient adapters. Zhang et al. \cite{zhang2025collm} further embed collaborative user and item representations directly into prompts in CoLLM to improve personalisation. Further research explores prompt tuning \cite{lester2021power}, prefix tuning \cite{li2021prefix}, and chain-of-thought prompting \cite{wei2022chain} to help LLMs reason over complex user histories.

\subsection{Inversion Attacks on LLMs}
A range of works has shown that sensitive inputs can be reconstructed from model outputs. Fredrikson et al. \cite{fredrikson2015model} first demonstrated model inversion by recovering private attributes from confidence scores. Liu et al. \cite{liu2023hqa} propose HQA Attack, a hard label black-box method that perturbs text inputs without gradient access, highlighting the fragility of language models even under minimal feedback. 
Membership-inference attacks reveal whether specific records were used for training through repeated softmax queries \cite{shokri2017membership}, and Carlini et al. \cite{carlini2021extracting} extract memorized training data from large language models. Unlike prior work focusing on memorized snippets, our goal is to reconstruct prompts issued to an LLM-based recommendation model, which contain highly sensitive preference and profile information. Foundational logit recovery techniques, such as iterative logit to text optimisation \cite{morris2023text} and multiple query statistical aggregation, enable \texttt{vec2text} style frameworks that refine textual reconstructions.

\subsection{Attacks on RecSys}
Existing research has extensively studied attacks against both traditional and deep-learning-based RecSys. These attacks primarily focus on undermining recommendation quality through various manipulation strategies: Burke et al. \cite{burke2005limited} introduce shilling attacks that inject fake ratings into collaborative filtering systems, and Christakopoulou \& Banerjee \cite{christakopoulou2019} show that small perturbations to user histories could mislead recommenders. 
Modern approaches employ sophisticated techniques such as reinforcement learning for poisoning attacks \cite{song2020poisonrec}, knowledge graph exploitation for targeted attacks \cite{song2022}, and social graph utilization for untargeted performance degradation \cite{fan2023untargeted}.
While these studies comprehensively address robustness challenges, none investigate the unique privacy risks inherent in LLM-empowered RecSys, a gap our work specifically addresses.
\section{CONCLUSION}

This study systematically reveals the privacy vulnerabilities inherent in LLM-empowered RecSys, highlighting their susceptibility to inversion attacks. 
By proposing an optimized inversion framework that integrates a \texttt{vec2text} engine with similarity-guided refinement, we demonstrate the feasibility of reconstructing user prompts including sensitive attributes and preference histories from model outputs with high fidelity, as validated through extensive experiments across two domains and two representative recommendation models.
Furthermore, we investigate the critical factors affecting inversion performance, such as domain alignment and prompt complexity. Collectively, these findings not only expose a previously underexplored privacy threat in modern recommendation pipelines, but also call for urgent attention to defensive strategies that mitigate the risks of prompt inversion in LLM-empowered RecSys.

\begin{acks}
This work is generously supported by the Australian Research Council under DECRA Grant No. DE250100032. We are also grateful to the anonymous reviewers for their insightful feedback.
\end{acks}

\bibliographystyle{ACM-Reference-Format}
\bibliography{software}

\end{document}